\documentclass[prd,article,twocolumn,preprintnumbers,superscriptaddress]{revtex4}
\usepackage{graphicx,amsfonts,color,comment,amsmath,hyperref,float}
\usepackage{amssymb}	

\usepackage{mathrsfs,amssymb}  
\usepackage{cancel}
\usepackage[normalem]{ulem}

\newcommand{\be}{\begin{equation}}
\newcommand{\ee}{\end{equation}}
\newcommand{\bea}{\begin{eqnarray}}
\newcommand{\eea}{\end{eqnarray}}

\begin{document}

\title{High-energy cosmic neutrinos as a probe of the vector mediator scenario in light of the muon $g-2$ anomaly and Hubble tension}

\author{Jose Alonso Carpio}
\affiliation{Department of Physics; Department of Astronomy and Astrophysics; Center for Multimessenger Astrophysics, The Pennsylvania State University, University Park, Pennsylvania 16802, USA}
\author{Kohta Murase}
\affiliation{Department of Physics; Department of Astronomy and Astrophysics; Center for Multimessenger Astrophysics, The Pennsylvania State University, University Park, Pennsylvania 16802, USA}
\affiliation{School of Natural Sciences, Institute for Advanced Study, Princeton, New Jersey 08540, USA}
\affiliation{Center for Gravitational Physics and Quantum Information, Yukawa Institute for Theoretical Physics, Kyoto, Kyoto 16802, Japan}
\author{Ian M. Shoemaker}
\affiliation{Center for Neutrino Physics, Department of Physics, Virginia Tech, Blacksburg, Virginia 24061, USA}
\author{Zahra Tabrizi}
\affiliation{Center for Neutrino Physics, Department of Physics, Virginia Tech, Blacksburg, Virginia 24061, USA}
\affiliation{Northwestern University, Department of Physics \& Astronomy, 2145 Sheridan Road, Evanston, IL 60208, USA}

\date{\today}
\begin{abstract}
We show that Gen-2 can probe
\end{abstract}
\preprint{}

\begin{abstract}
In light of the recent Muon $g-2$ experiment data from Fermilab, we investigate the implications of a gauged $L_{\mu} - L_{\tau}$ model for high energy neutrino telescopes. 
It has been suggested that a new gauge boson at the MeV scale can both account for the Muon $g-2$ data and alleviate the tension in the Hubble parameter measurements. 
It also strikes signals at IceCube from the predicted resonance scattering between high-energy neutrinos and the cosmic neutrino background. We revisit this model based on the latest IceCube shower data, and perform a four-parameter fit to find a preferred region. 
We do not find evidence for secret interactions. The best-fit points of $m_{Z'}$ and $g_{\mu\tau}$ are $\sim10$~MeV and $\sim0.1$, respectively, depending on assumptions regarding the absolute neutrino masses, and the secret interaction parameter space allowed by the observed IceCube data overlaps with the regions of the parameter space that can explain the muon $g-2$ anomaly and Hubble tension as well.
We demonstrate that future neutrino telescopes such as IceCube-Gen2 can probe this unique parameter space, and point out that successful measurements would infer the neutrino mass with $0.06~{\rm eV}\lesssim \Sigma m_\nu\lesssim 0.3~{\rm eV}$.
\end{abstract}


\maketitle

\section{Introduction}
The recent data from the Fermilab Muon $g-2$ Collaboration indicates that muon magnetic moment may disagree with phenomenological predictions from the Standard Model (SM)~\cite{Abi:2021gix} consistent with the earlier E821 experiment at Brookhaven~\cite{Bennett:2006fi}. Although it could be explained by the SM physics through the hadronic vacuum polarization~\citep{Davier:2010nc,Davier:2017zfy,Davier:2019can,Borsanyi:2020mff}, this may indicate beyond the Standard Model (BSM) physics coupled to the muons. As a consequence of electroweak gauge symmetry, modifications to muon physics would imply modified neutrino physics as well given that charged leptons and neutrinos come together in $SU(2)$ doublets. In this paper, we explore an example of this in the context of gauged lepton number. A very large number of possible interpretations of the new Muon g-2 results have already appeared, including supersymmetry \cite{Zhang:2021gun,Baum:2021qzx,Ahmed:2021htr,Cox:2021nbo,Abdughani:2021pdc,VanBeekveld:2021tgn,Yin:2021mls,Han:2021ify,Buen-Abad:2021fwq,Wang:2021bcx,Cao:2021tuh,Gu:2021mjd,Athron:2021iuf,Aboubrahim:2021xfi,Yang:2021duj} new $U(1)$ gauge symmetries \cite{Greljo:2021xmg,Zu:2021odn,Cadeddu:2021dqx,Yang:2021duj,Kawamura:2021ygg,Li:2021lnz,Borah:2021jzu,Buras:2021btx}, dark matter \cite{Arcadi:2021cwg,Bai:2021bau,Zu:2021odn,Lu:2021vcp} axions and axion-like particles \cite{Ge:2021cjz,Brdar:2021pla}, Higgs doublet models \cite{Ferreira:2021gke,Wang:2021fkn,Chen:2021vzk,Athron:2021iuf,Chun:2021rtk,Arcadi:2021yyr,Li:2021lnz}, 331 models \cite{Li:2021poy}, seesaw models \cite{Escribano:2021css} and leptoquarks \cite{ColuccioLeskow:2016dox,Crivellin:2020tsz}

As is well known, gauging the lepton number combination $L_{\mu}-L_{\tau}$ is anomaly free~\cite{He:1990pn,He:1991qd}. It is also experimentally challenging to probe given that its main effects are to modify the interactions of unstable charged leptons and neutrinos. Intriguingly, this combination of gauged lepton numbers can both explain $(g-2)_{\mu}$ and be consistent with the constraints from null experiments~\cite{Altmannshofer:2014cfa,Altmannshofer:2014pba,Bauer:2018onh}. In principle, this scenario can be tested at NA64$\mu$, the European Spallation Source,  DARWIN~\cite{Amaral:2021rzw}, the Missing Muon Momentum (${\rm M}^{3}$) experiment at Fermilab~\cite{Kahn:2018cqs}, and the High Luminosity-LHC~\cite{Galon:2019owl}.

\begin{figure}[b!]
\includegraphics[angle=0,width=.45\textwidth]{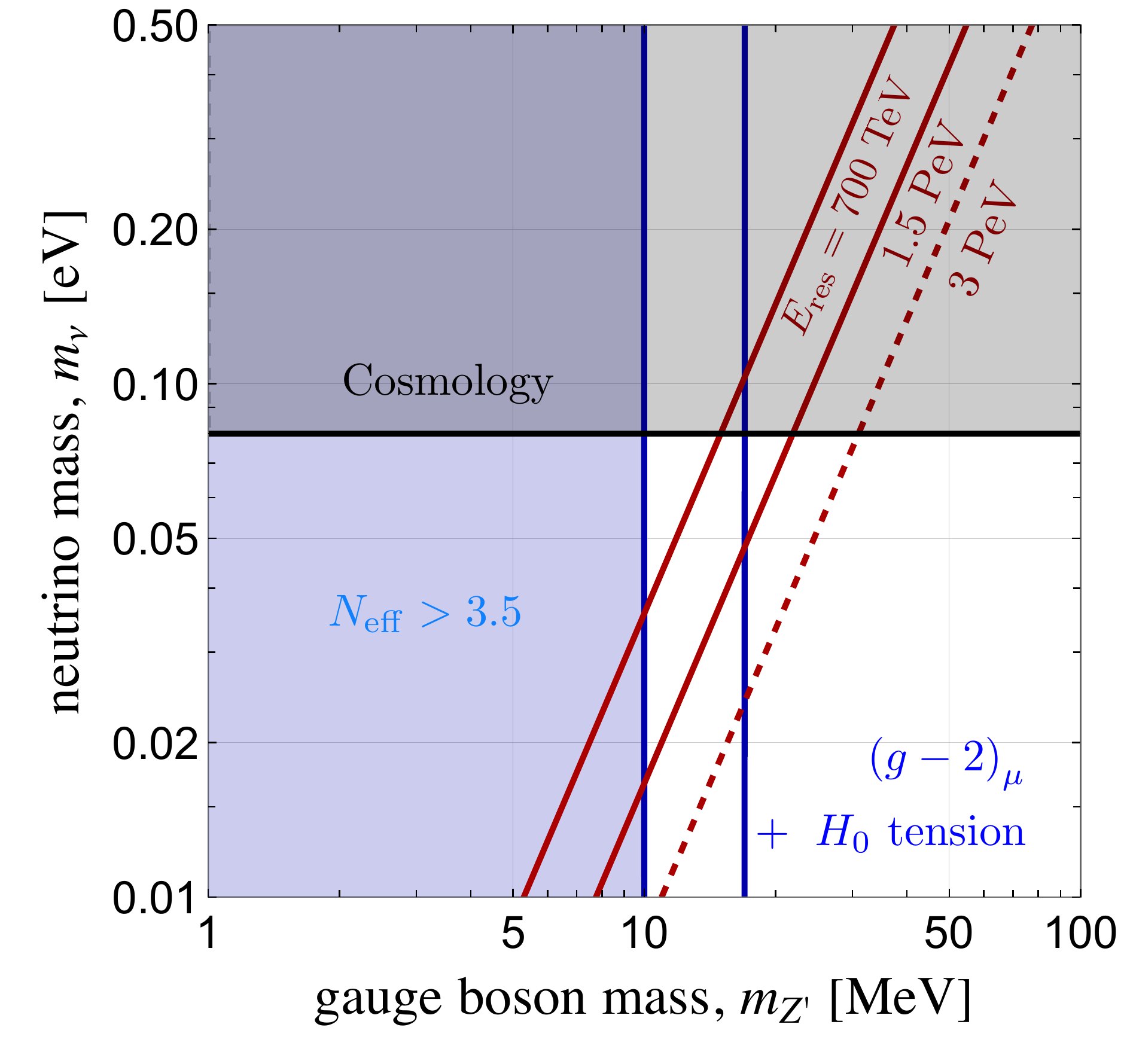}
\caption{Preferred and excluded regions in the mediator and neutrino mass plane. The part of the parameter space, which is allowed by IceCube data, is shown in the red curves, while the region between the blue curves is the part allowed by the combination of muon $(g-2)$ and Hubble tension. The combination of data sets can be used to infer non-trivial bounds on the absolute neutrino masses (see text for details). 
}
\label{fig:params}
\end{figure}


This work focuses on a different probe of gauged $L_{\mu} - L_{\tau}$, involving only neutrinos. We study the current and future sensitivity of the IceCube neutrino telescope to such new gauge interactions, by examining in detailed the modifications to the spectrum of high-energy cosmic events~\cite{Ioka:2014kca,Ng:2014pca}. 
Such interactions can also alleviate the tension in Hubble parameters between the local value and cosmic microwave background (CMB) data, through delaying the neutrino free-streaming by self-interactions~\cite{Cyr-Racine:2013jua,Archidiacono:2013dua,Lancaster:2017ksf,Oldengott:2017fhy,Kreisch:2019yzn,Blinov:2019gcj} or adding to the effective number of relativistic (neutrino) species~\cite{Weinberg:2013kea,Berlin:2018ztp,DEramo:2018vss,Escudero:2019gzq}.
We find that the data from IceCube and the Muon $g-2$ collaboration can be combined to yield a non-trivial determination of neutrino masses. Earlier work has also examined the impact of gauged $L_{\mu} - L_{\tau}$  at IceCube~\cite{Araki:2014ona,Kamada:2015era,DiFranzo:2015qea}. We also note that extended gauge symmetries (e.g., different baryon and lepton number combinations) may allow one to also connect to the LMA-Dark solution of neutrino oscillations for gauge boson masses in the range we are considering here~\cite{Farzan:2015doa,Denton:2018xmq}. We note that Ref.~\cite{Okada:2019sbb} commented on the connections between IceCube, Hubble tension, and muon $(g-2)$ in the $L_{\mu} - L_{\tau}$ model. 

\begin{figure}[t!]
\includegraphics[angle=0,width=0.48\textwidth]{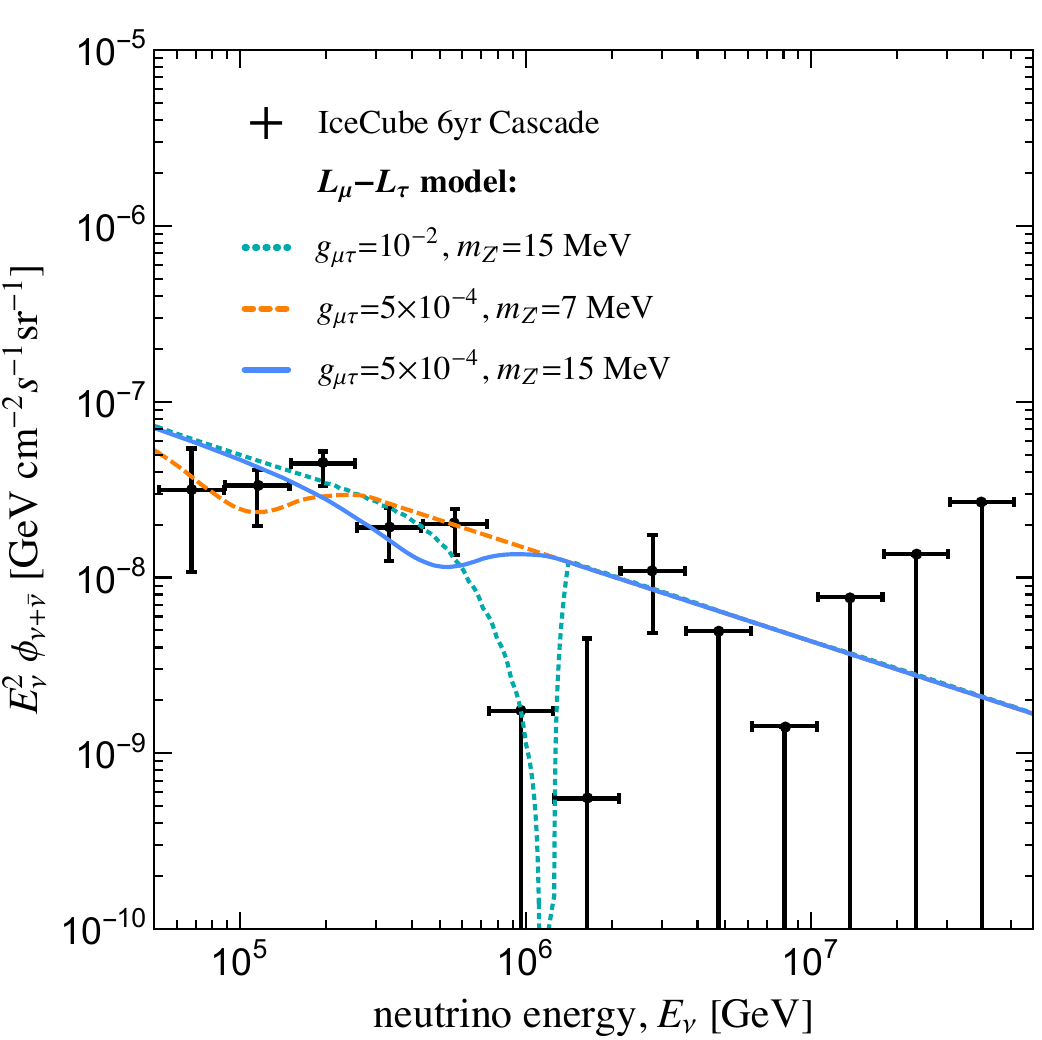}
\caption{The astrophysical fluxes of neutrinos per neutrino flavor for the IceCube 6-year shower events (black crosses)~\cite{Aartsen:2020aqd} and the $L_\mu-L_\tau$ model (shown with colored curves). For the latter we have fixed the astrophysical parameters to the IceCube best fit values $(\Phi_{\rm WB},s_\nu)=(1.66,2.53)$.}  
\label{fig:fluxNP}
\end{figure}

\begin{figure*}[t!]
\includegraphics[angle=0,width=1\textwidth]{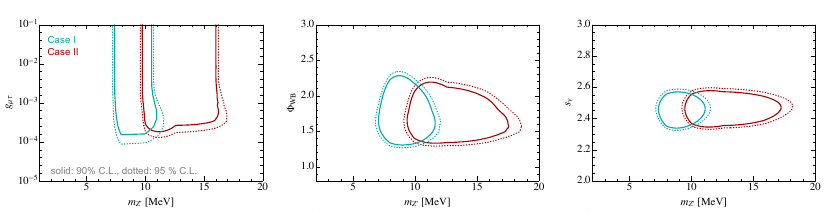}
\caption{Best-fit regions at 90$\%$ CL (solid) and 95$\%$ CL (dotted). Here we fit to a four parameter model, in which the astrophysical neutrino flux is parametrized by the spectral index and normalization, $(s_{\nu},\Phi_{\rm WB})$, while the particle physics of the $U(1)_{L_{\mu}-L_{\tau}}$ model is fixed by the parameters $(m_{Z'},g_{\mu \tau})$. Cases I and II refer to two different possibilities for the spectrum of neutrino mass eigenstates (see text for details). 
}
\label{fig:best}
\end{figure*}

In Fig.~\ref{fig:params} we show the IceCube preferred region in red in the plane of the gauge boson mass and the neutrino mass. 
Here the resonance energy in the observer frame is $E_{\rm res}=m_{Z'}^2/[2m_\nu(1+\bar{z})]$, where $\bar{z}\sim1$ is the typical redshift of the neutrino sources.
The gray shaded area shows the region of neutrino masses excluded by cosmology. Given that none of the individual neutrino masses can exceed the cosmological bound on the sum of neutrino masses, we display the Planck 2018 bound $\sum m_{\nu} < 0.24$~\cite{Planck:2018vyg}. Notice that the region between $m_{Z'}\sim10-17$~MeV shows the range of the gauge boson mass that may explain the $(g-2)_{\mu}$ observations while remaining consistent with the null results from CCFR~\cite{Altmannshofer:2014pba,Mishra:1991bv} and Borexino~\cite{Kamada:2015era,Harnik:2012ni,Bellini:2011rx}. For simplicity we have assumed the natural level of loop-induced kinetic mixing for the Borexino constraint, but in principle this can be relaxed by allowing model-dependent additional particles in the loop. 
This would only allow for slightly lighter gauge boson masses. Similar masses and gauge couplings can also alleviate the Hubble tension via the extra contribution to the radiation density from the light vector particle (e.g., Ref.~\cite{Escudero:2019gzq}). 

The remainder of this paper is as follows. In the next section, we introduce the model and the neutrino-neutrino cross section the model predicts. In Sec.~III we consider implications of the present IceCube shower data can provide, being careful to allow for fairly weak priors on both particle and astrophysical parameters. We discuss the potential of the next generation detectors in confirming or excluding the model in Sec.~IV and we conclude in Sec.~V.

\section{Gauged $L_{\mu}-L_{\tau}$ Model}
We consider a model of the gauged $L_{\mu}-L_{\tau}$ number~\cite{He:1990pn,He:1991qd}, with the Lagrangian %
\be
\mathscr{L} \supset g_{\mu \tau} j_{\mu-\tau}^{\alpha} Z'_{\alpha}- \frac{m_{Z'}^{2}}{2} Z'_{\alpha} Z'^{\alpha},
\ee
where $g_{\mu \tau}$ is the gauge coupling, $Z'$ is the new gauge boson with mass $m_{Z'}$, and the current associated with the new symmetry is 
\be
j_{\mu-\tau} \equiv \bar{L}_{2} \gamma_{\alpha} L_{2} + \bar{\mu}_{R} \gamma_{\alpha} \mu_{R} - \bar{L}_{3} \gamma_{\alpha} L_{3}- \bar{\tau}_{R} \gamma_{\alpha} \tau_{R},
\ee
where $L_{i}$ is the lepton doublet of the $i^{th}$ generation. 
This new gauge interaction allows for high-energy neutrinos to scatter on the neutrinos of the cosmic neutrino background (C$\nu$B). The most significant effect is the $s$-channel scattering cross section, which in terms of mass eigenstates can be written as, 
\be
\sigma( \nu_{i} \nu_{j} \rightarrow \nu \nu) = \frac{2}{3\pi} g_{\mu\tau}^{4}Q_{ij}^{2}~\frac{s_j}{(s_j-m_{Z'}^{2})^{2} + m_{Z'}^{2} \Gamma_{Z'}^{2}}, 
\ee
where for a given incoming neutrino energy $E_{\nu}$ the Mandelstam variable $s_j$ is $s_j \approx 2 m_j E_{\nu}$, where $\{m_1,m_2,m_3\}$ are the masses of the mostly active neutrinos, and the width is $\Gamma_{Z'} = g_{\mu \tau}^{2} m_{Z'}/(12 \pi)$. We have also defined the effective charge $Q_{ij}$ in the above for scattering of the mass eigenstates:  $Q_{ij} = \left( U^{\dagger} G U \right)_{ij}$, where $G = {\rm diag} (0,+1,-1)$, and $U$ is the Pontecorvo–Maki–Nakagawa–Sakata (PMNS) matrix. In principle, the $t$-channel contributions can also be relevant at large couplings, e.g. $g\gtrsim 0.1$~\cite{Blum:2014ewa,Cherry:2014xra,Cherry:2016jol}, which is out of the range we consider in this work, and we have checked numerically that we can neglect it here.

The neutrinos scattering off each other can cause the depletion of astrophysical neutrinos at the resonant energies $E_j= m^2_{Z'}/(2m_j)$~\cite{Ioka:2014kca,Ng:2014pca,Ibe:2014pja}. Because of $\Gamma_{Z'} \ll m_{Z'}$, the cross section would be localized around $E_\nu=E_j$: $\sigma(\nu_{i} \nu_{j} \rightarrow \nu \nu)=\sigma_{ij}^R E_\nu\delta(E_\nu-E_j)$ where $\sigma_{ij}^R=\frac{2g_{\mu\tau}^{4}Q_{ij}^{2}}{3m_{Z'}\Gamma_{Z'}} $ is the effective cross section, averaged over the resonance width. We can calculate the neutrino optical depth as~\cite{Creque-Sarbinowski:2020qhz}:
\be
\tau_i(E_\nu,t',t)=\sum_j \tau^{R}_{ij}(E_\nu,t)\Theta(z_j(E_\nu,t)-z)\Theta(z'-z_j(E_\nu,t)),
\ee
using
$
\tau^{R}_{ij}=\frac{\Gamma_{ij}^R}{H(z_j)}\frac{1+z}{1+z_j}$,
where $\Gamma_{ij}^R(t)=n_j(t)\sigma_{ij}^R$,  $z_j=(1+z)E_j/E_\nu-1$ and $n(t)$ is the neutrino number density. Here $H(z)$ is the Hubble parameter at redshift $z$ and $\Theta(x)$ is the Heaviside function. We can calculate the flux of BSM-mediated astrophysical neutrinos $\nu_i$ as (e.g.,~\cite{Murase:2015xka})
\bea
\Phi_i(E_\nu)&=&\frac{c}{4\pi}\int dz \frac{1}{H(z)}R(z)\frac{dN_\nu}{dE'_\nu}e^{-\tau_i(E_\nu,z)}
\eea
where $R(z)(dN_\nu/dE'_\nu)$ is the differential rate density of the astrophysical neutrinos.
For the redshift evolution of sources, $R(z)$, we assume that they are distributed according to the star-formation rate~\cite{Yuksel:2008cu}.
We do not consider effects of cascades as an approximation, which is reasonable because its effect is a factor of 2 for $s_\nu\sim2.5$ we consider~\cite{Blum:2014ewa}. 

Finally, because the neutrino decoherence time scale is much smaller than other relevant time scales, the fluxes of the neutrino flavors $\nu_\alpha$ are given by $\Phi_{\alpha}=\sum_i |U_{\alpha i}|^2\Phi_i$.

\section{Implications of High-Energy Neutrino Data}
Secret neutrino self-interactions lead to striking spectral distortions at IceCube~\cite{Ioka:2014kca,Ng:2014pca,Ibe:2014pja,Blum:2014ewa,Araki:2014ona}. 
To find this spectral feature, it is crucial to use a data sample with a good energy resolution. 
High-energy starting events (HESEs), including showers and starting tracks, are often used, in which the deposit energy distribution is calculated~\cite{Blum:2014ewa,Shoemaker:2015qul}. 
For this purpose, for our numerical analysis we use the 6-yr shower data sample of IceCube, which is dominated by the electron and tau contributions to the data~\cite{Aartsen:2020aqd}. 
This has an advantage of having more events especially at lower energies, which enables us to determine the spectral index $s_\nu$ (defined below) better. (Note that the 6-yr shower analysis result, $s_\nu=2.53\pm0.07$, is consistent with the 7.5~yr HESE analysis result, $s_\nu=2.87_{-0.19}^{+0.20}$~\cite{Abbasi:2020jmh}, within $\sim1.5\sigma$).  
The ordinary SM fit that the IceCube collaboration performs, fits to an unbroken power-law with the slope $s_\nu$
\bea
\Phi(E_\nu) &=& 3\times 10^{-18}~\left({\rm GeV}\cdot {\rm s} \cdot {\rm cm}^{2} \cdot {\rm sr} \right)^{-1}\nonumber\\
&\times&\Phi_{\rm WB} \left(\frac{E_{\nu}}{100~{\rm TeV}}\right)^{-s_\nu}
\eea
where $\Phi(E_\nu)(100~{\rm TeV})\equiv3\times10^{-18}\left({\rm GeV}\cdot {\rm s} \cdot {\rm cm}^{2} \cdot {\rm sr} \right)^{-1}$\\$\Phi_{\rm WB}$ is the all-flavor flux at 100~TeV. The IceCube collaboration officially found the best-fit values, $s_\nu=2.53\pm0.07$ and $\Phi_{\rm WB}=1.66 ^{+ 0.25}_{-0.27}$. 

To perform our analysis, instead of calculating the expected number of events we calculate $E_\nu^2\Phi(E_\nu)$ at each bin of energy and compare it with the flux data given in Fig.~3 of Ref.~\cite{Aartsen:2020aqd}. We define the following log-likelihood function:
\bea 
\label{eq:chisq}
\chi^2_{\rm spectral}
&=&2\mathcal{N}\sum_{j}\Bigg\{\Big((E_\nu^2\Phi)_j^{\rm{Th}}-(E_\nu^2\Phi)_j^{\rm{Cascade}}\Big)\\&-&(E_\nu^2\Phi)_j^{\rm{Cascade}}\log\Big(\frac{(E_\nu^2\Phi)_j^{\rm{Th}}}{(E_\nu^2\Phi)_j^{\rm{Cascade}}}\Big)\Bigg\}\,,\nonumber
\eea
where $(E_\nu^2\Phi)_j^{\rm{Cascade}}$ are the IceCube cascade fluxes at each bin of energy while $(E_\nu^2\Phi)_j^{\rm{Th}}$ is the theory prediction. We consider the neutrino energies $4.7~{\rm{TeV}} \leq E_\nu \leq 9.1\times10^4~{\rm{TeV}}$, sorted into 13 log-spaced bins. The parameter $\mathcal{N}$ is a normalization factor that contains information on the effective volume, cross section, and observation time, which guarantees each term has the same magnitude as the observed number of signal events at IceCube. We have checked that by using $\mathcal{N}=3\times10^8$ we can reproduce the IceCube results, getting the best fit values $s_\nu=2.49$ and $\Phi_{\rm WB}=1.65$ with an allowed region similar to Fig.~2 of Ref.~\cite{Aartsen:2020aqd}. We caution that our approach is only approximate. At present, detailed information on the event selection is not publicly available for the shower data, and we do not take into account details such as effects of the energy smearing due to neutral-current interactions and systematic errors from the atmospheric background. 
Nevertheless, we confirm that our analyses are broadly consistent with the IceCube results, so the method is accurate enough for the purpose of this work.

In Fig.~\ref{fig:best} we display the marginalized best-fit regions. In each panel, the two parameters not shown have been marginalized over. We consider two cases for the neutrino mass spectrum. Case I fixes the neutrino masses to   
$
(m_{1},m_{2},m_{3}) = (0.03,0.031,0.059)~{\rm eV}.
$
%
While in case II for which the masses are nearly degenerate we take
$
(m_{1},m_{2},m_{3}) = (0.0871,0.0876,0.1)~{\rm eV}.
$
%

To fit our model to the IceCube data we assume that $R(z)$ follows the star-formation rate, which is reasonable for most astrophysical models~\cite{Murase:2016gly}. Following the results of Ref.~\cite{DiFranzo:2015qea} we have concluded that different $R(z)$ models do not change our results significantly, and the effect on the shape is negligible and very hard to observe, see e.g., Fig.~10 of~\cite{Murase:2007yt}. For $g_{\mu\tau}\neq 0$, the observed flux is no longer a power-law, so the parameter $\Phi_{\rm WB}$ used in our fits will be defined so as to normalize the flux at 100 TeV under the standard case $g_{\mu\tau}=0$: $\Phi_\alpha|_{g_{\mu\tau}=0}(100~{\rm TeV})/\Phi_{\rm WB}\equiv3\times10^{-18}\left({\rm GeV}\cdot {\rm s} \cdot {\rm cm}^{2} \cdot {\rm sr} \right)^{-1}$. We fit the data to both $(\Phi_{\rm WB},s_\nu)$ as the collaboration does, but also the two new particle physics parameters associated with the new gauge symmetry, $g_{\mu\tau}$ and $m_{Z'}$. We find the best-fit points of $m_{Z'}=7.94~$MeV and $g_{\mu\tau}=0.10$ for Case I and $m_{Z'}=11.27~$MeV and $g_{\mu\tau}=0.09$ for Case II.
Yet, the secret interaction model does not significantly improve the fit to the IceCube data (the difference between the two models are less than $1\sigma$ C.L.).
\begin{figure}[t!]
\includegraphics[angle=0,width=.48\textwidth]{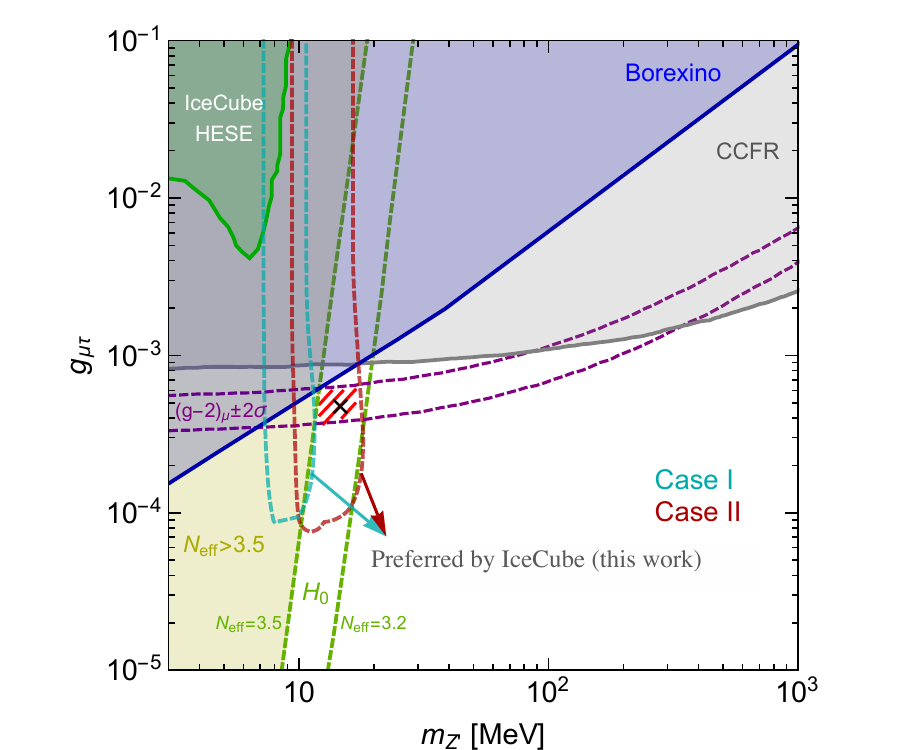}
\caption{Constraints and preferred regions for the gauged $L_{\mu}-L_{\tau}$ model. 
The shaded regions are constrained by the current experiments: the gray region is excluded by the trident measurement at the CCFR experiment \cite{Mishra:1991bv}, the blue region is excluded by Borexino \cite{Bellini:2011rx}, the green region is a limit from the IceCube HESE data \cite{Bustamante:2020mep} and the yellow region is bounded by cosmology \cite{Escudero:2019gzq}. 
The dashed curves are the preferred regions that explain or alleviate the anomalies: the purple band is the region favored by the $(g-2)_\mu$ discrepancy, the green band is the region which alleviates the tension in the Hubble parameter measurements \cite{Escudero:2019gzq}, the two cyan and red regions represent the two cases considered in this work as regions preferred by the IceCube shower data.}
\label{fig:moneyplot}
\end{figure}


Let us first discuss the 
left panel, which displays the best-fit region in the mass-coupling plane. 
We see that both Cases I and II prefer a relatively narrow range of vector masses at $90\%$ CL (solid) and $95\%$ CL (dotted). This fits the expectation from the $s$-channel resonance cross section which is sharply peaked around
\be
E_j^{{\rm res}} = \frac{m_{Z'}^{2}}{2 m_{j}} \simeq 0.5~{\rm PeV}~\left(\frac{m_{Z'}}{5~{\rm MeV}}\right)^{2}~\left(\frac{0.06~{\rm eV}}{m_{j}}\right).
\ee
At present, the IceCube shower and HESE data lack statistics in the $0.2-1$~PeV range~\cite{Aartsen:2020aqd,Abbasi:2020jmh}. This possible dip-like feature has been paid attention to for several years~\cite{Araki:2014ona}. 
The self-interaction cross section around these resonance energies can induce significant depletion of neutrino flux.

The fact that there appears to be no upper bound on the coupling in the 
left panel of Fig.~\ref{fig:best} is because of the dip in the $0.2-1$~PeV region. 
Larger couplings result in greater flux depletion, as shown in Fig.~\ref{fig:fluxNP}. 
Let us also mention that depending on how exactly we calculate the contained energy information in each bin there could be less than $20\%$ of analysis related uncertainties which could result in moving the mass regions slightly to the left. 
The conclusions of this work are however unchanged. 

We show in Fig.~\ref{fig:moneyplot} the preferred regions by IceCube we find in this work in the mass-coupling plane, accompanied by the excluded region by the CCFR experiment 
\cite{Mishra:1991bv} (gray shaded region), the blue shaded region is the excluded region by Borexino \cite{Bellini:2011rx}, while the purple band represents the preferred $2\sigma$ region from the $(g-2)_\mu$ discrepancy~\cite{Amaral:2021rzw}. It is also important to note that such leptophilic interactions mentioned in this work can also affect the relativistic degrees of freedom of neutrinos and so to avoid tension with cosmology it requires that $m_{Z'}\gtrsim 10$~MeV so that $\Delta N_{\rm eff}<0.5$ ~\cite{Escudero:2019gzq}. We show the excluded region in yellow. It was also mentioned in Ref.~\cite{Escudero:2019gzq} that an additional $Z'$ boson can also alleviate the Hubble tension (even though it cannot be fully resolved), for the mass-coupling region shown with the green band. Last but not least, one could see all the favored regions cross each other at $m_{Z'}=10-17$~MeV and $g_{\mu\tau}=(4-6.5)\times10^{-4}$, shown in the hashed red region. Note that the cosmological limit used here considers the kinetic mixing, which is stronger than limits only with neutrino self-interactions although it depends on $\Delta N_{\rm eff}$~\cite{Aarssen:2012fx,Ahlgren:2013wba,Kamada:2015era}.   

The preferred regions are intriguing because results obtained from three independent measurements meet each other. 
On the other hand, we stress that the IceCube data have not shown evidence for secret neutrino self-interactions, by which we can place an upper limit on the coupling rather than the preferred region. We also show the previous results by Ref.~\cite{Bustamante:2020mep} in Fig.~\ref{fig:moneyplot}. The constraints are weaker than the limits from Borexino as well as other laboratory experiments such as the kaon-decay measurement implying $g_{\mu\tau}\lesssim0.01$~\cite{Blum:2014ewa}. 

\section{Future Prospects}\label{Sec:future}
%
\begin{figure}[t!]
\includegraphics[angle=0,width=.45\textwidth]{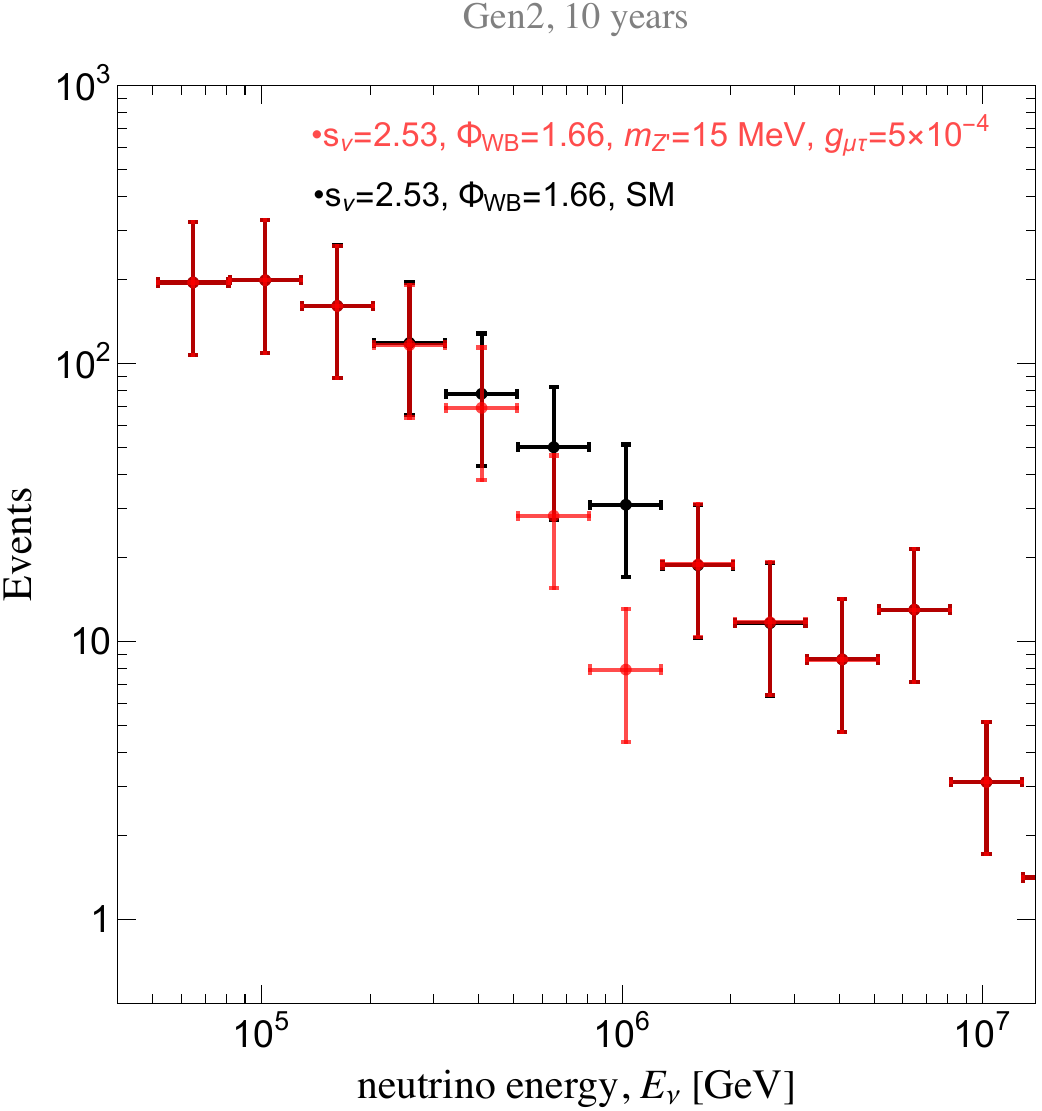}
\caption{Neutrino energy distribution of mock events expected in 10 years of running with IceCube-Gen2, using the SM best-fit points for the spectral index and the normalization. The red data points show the mock data with secret interactions for $m_{Z'}=15$~MeV and $g_{\mu\tau}=5\times10^{-4}$, which deviate by $\sim5\sigma$ from the black ones corresponding to the SM scenario.}
\label{fig:gen2}
\end{figure}


Lastly, we consider the impact of next-generation detectors such as IceCube-Gen2~\cite{Aartsen:2020fgd} on the gauged $L_{\mu}-L_{\tau}$ scenario considered in this work. 
In particular, we are interested in the unique parameter space, where the muon $g-2$ anomaly solution, Hubble tension alleviation, and IceCube preferred region are overlapping. For demonstration, we adopt $m_Z'=15$~MeV and  $g_{\mu\tau}=5\times10^{-4}$ as the fiducial scenario (shown as the black cross in Fig.~3).
Using the zenith-angle-averaged effective areas for shower-type events based on Fig.~25 from Ref.~\cite{Aartsen:2020fgd} we estimate the number of events coming from a given neutrino flux. As in the analysis in the previous section, this approach here is different from those in Refs.~\cite{Blum:2014ewa,Shoemaker:2015qul} that used the energy deposited in the detector. 
In Fig.~\ref{fig:gen2} we compare neutrino spectra with and without BSM neutrino-neutrino scatterings in red and black data points respectively, assuming 10 years of IceCube-Gen2 data and the neutrino spectrum with $s_\nu=2.53$ and $\Phi_{\rm WB}=1.66$. 
It shows that with statistics expected in IceCube-Gen2, the dip feature will be evident if it exists. We also compute the resulting $\chi^2$, and find that our fiducial scenario would be $\sim 5\sigma$ discrepant with the SM case without secret interactions. Although results depend on our understanding of astrophysical components, this demonstrates that such model can be probed by the IceCube telescopes.   

In Fig.~\ref{fig:gen2}, only statistical errors are considered. In reality, there are other systematics, which need to be taken into account. As noted above, the deposited energy is smaller than the neutrino energy, which can make the expected dip broader. The atmospheric background gives additional systematics in the analysis. On the other hand, this analysis only used the shower data. Muon track data including starting and through-going events should also give us information. One may be able to further uncover the nature of the preferred model of secret self-interactions by combining spectral and flavor modifications~\cite{Shoemaker:2015qul,Barenboim:2019tux}, and global analyses as in Ref.~\cite{Aartsen:2015ita} will be more powerful.

\section{Summary and Discussion}\label{Sec:Conc}
It has been suggested that the gauged $L_\mu-L_\tau$ model accounts for the muon $g-2$ anomaly. 
High-energy neutrino data provide an independent test for this model through dip signatures caused by secret neutrino self-interactions. We showed in this work that the current 6-yr shower data of IceCube prefers couplings and masses consistent with the Muon $g-2$ data, which also overlaps with the parameter space alleviating the Hubble tension~\cite{Escudero:2019gzq}. 

We have performed a likelihood analysis similar to ref.~\cite{IceCube:2020phf}, where we show our results based on the assumption that the gauged $L_{\mu}-L_{\tau}$ model is ``preferred'' over the null hypothesis. This is because the current IceCube shower data have a paucity in the $0.2-1$~PeV range, but the sensitivity (or constraint) is weaker as indicated by the HESE constraint in Fig.~\ref{fig:moneyplot}. 
Gen2 can be sensitive to the preferred region hinted by ICeCube. 

Future neutrino experiments such as IceCube-Gen2 will be sensitive to the parameter space indicated by Fig.~\ref{fig:gen2}, and may confirm the dip. But one should keep in mind that the dip can also be caused by astrophysical sources. For example, this may reflect two or more astrophysical populations~\cite{Chen:2014gxa,Murase:2015xka}. The dip could also be caused by the Bethe-Heitler process that can be important between $pp$ and $p\gamma$ interactions and/or the combination of multi-pion production and pileup due to the meson/muon cooling. However, these details are model dependent, and it is beyond the scope of this work to perform the BSM analysis taking account of such astrophysical systematics.

If this result is confirmed, the combination of the data may not only reveal the existence of a new fundamental symmetry, but also uncover the neutrino mass spectrum. We found that the required parameter space is narrow, and the total neutrino mass has to range from 0.06~eV to 0.3~eV. This is also encouraging for future neutrino mass measurements (e.g.,~\cite{Abazajian:2016yjj,Dvorkin:2019jgs,Betti:2019ouf,Esfahani:2017dmu}). 

We also demonstrated that the gauged $L_\mu-L_\tau$ scenario for the muon $g-2$ anomaly and Hubble tension alleviation can critically be tested by IceCube-Gen2. This result is consistent with the previous work that showed next-generation neutrino telescopes can reach the limit expected in the mean free path limit~\cite{Shoemaker:2015qul}. 

Another important test with high-energy neutrinos is to utilize multimessenger observations from individual neutrino sources~\cite{Kelly:2018tyg,Murase:2019xqi,Koren:2019wwi}. In particular, BSM neutrino echoes -- delayed neutrino emission through secret neutrino-neutrino scatterings provide a test that is insensitive to the unknown astrophysical spectrum. Ref.~\cite{Murase:2019xqi} showed that IceCube-Gen2 can reach $g_{\mu\tau}\sim{10}^{-4}-{10}^{-2}$ for the vector mediator scenario.

\begin{acknowledgements}
The work of K.M. is supported by the NSF Grant No.~AST-1908689, No.~AST-2108466 and No.~AST-2108467, and KAKENHI No.~20H01901 and No.~20H05852. The work of I.M.S. and Z.T. is supported by the U.S. Department of Energy under the award number DE-SC0020250. Z.T. appreciates the useful discussions with Joachim Kopp.\\
\end{acknowledgements}

\appendix
\begin{figure}[t!]
\includegraphics[angle=0,width=.45\textwidth]{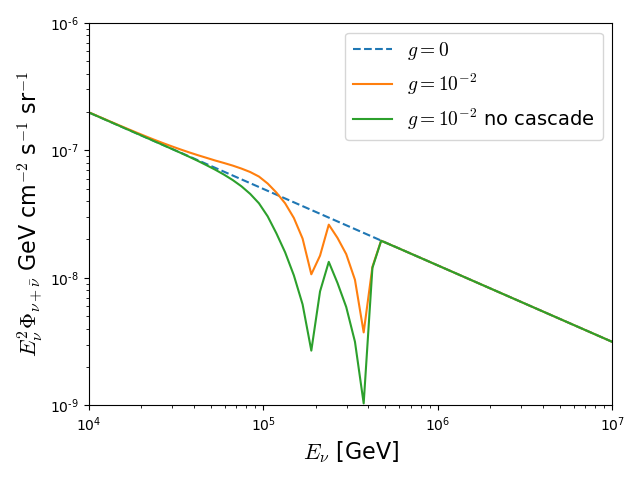}
\caption{Same as Fig.~\ref{fig:fluxNP}, but comparing the fluxes with up- scattering (orange) and without up- scattering (green) curves.}
\label{fig:mass}
\end{figure}

\section{Cascades Effects}
In order to confirm that the up- scattering of neutrinos does not affect the analysis of this work we have made Fig.~\ref{fig:mass}. To get the "no cascade" plot, we have collected all the particles that never scattered. As it should be expected the "no cascade" case is always below the cascade one, because in the latter we are still getting some energy deposited between resonances from scattered particles. However, the difference is about a factor of two at the peak located between resonances, and it does not significantly affect the preferred region we have obtained in this analysis.\\


\bibliography{kmurase.bib}

\end{document}